\documentclass{article}
\usepackage{amsmath,epsfig}
\usepackage{multirow}
\usepackage{url}
\usepackage{bm}
\usepackage{verbatim}
\usepackage{graphics,setspace}
\usepackage{graphicx}
\usepackage[preprint]{spconfa4}

\copyrightnotice{978-1-6654-3864-3/21/\$31.00 ©2021 IEEE}

\let\OLDthebibliography\thebibliography
\renewcommand\thebibliography[1]{
  \OLDthebibliography{#1}
  \setlength{\parskip}{0pt}
  \setlength{\itemsep}{0pt plus 0.3ex}
}

\makeatletter
\newcommand{\thickhline}{%
	\noalign {\ifnum 0=`}\fi \hrule height 1pt
	\futurelet \reserved@a \@xhline
}

\begin{document}\sloppy

\def\x{{\mathbf x}}
\def\L{{\cal L}}

\title{FastSVC: Fast Cross-Domain Singing Voice Conversion \\ with Feature-wise Linear Modulation}
%

\name{Songxiang Liu$^{1,*}$\thanks{*Work done during internship at Tencent AI Lab.}, Yuewen Cao$^{1,*}$, Na Hu$^2$, Dan Su$^2$, Helen Meng$^1$}

\address{
 $^1$Human-Computer Communications Laboratory,
 The Chinese University of Hong Kong\\
  $^2$Tencent AI Lab}

\maketitle

\begin{abstract}
This paper presents FastSVC, a light-weight cross-domain singing voice conversion (SVC) system, which can achieve high conversion performance, with inference speed 4x faster than real-time on CPUs.
FastSVC uses Conformer-based phoneme recognizer to extract singer-agnostic linguistic features from singing signals. A feature-wise linear modulation based generator is used to synthesize waveform directly from linguistic features, leveraging information from sine-excitation signals and loudness features. The waveform generator can be trained conveniently using a multi-resolution spectral loss and an adversarial loss.
Experimental results show that the proposed FastSVC system, compared with a computationally heavy baseline system, can achieve comparable conversion performance in some scenarios and significantly better conversion performance in other scenarios. Moreover, the proposed FastSVC system achieves desirable cross-lingual singing conversion performance. The inference speed of the FastSVC system is 3x and 70x faster than the baseline system on GPUs and CPUs, respectively.

\end{abstract}
\section{Introduction}
\label{sec:intro}

Human singing is an important way of information transmission, emotional expression and entertainment. Enabling machine the ability to produce high-fidelity singing voice can enrich the way of human-computer interaction.
This paper focuses on a singing synthesis related task, i.e., singing voice conversion (SVC), which aims at converting the voice of one singer to that of other singers without changing the underlying content and melody.

In terms of whether parallel singing datasets are used during training, which are composed of paired samples among singers singing the same content, current SVC approaches can be categorized into two classes: parallel SVC and non-parallel SVC. 
Most initial attempts for SVC belong to the parallel SVC class, which model parallel training samples using statistical methods, such as Gaussian mixture model (GMM) based many-to-many eigenvoice conversion \cite{toda2007one}, direct waveform modification based on spectrum difference \cite{kobayashi2014statistical,kobayashi2015statistical}. Artificial neural network (ANN) based approaches are also proposed to improve conversion performance \cite{hono2019singing,sisman2019singan}.

Since parallel singing datasets are expensive to collect in large scale, many non-parallel SVC approaches are proposed. WaveNet \cite{oord2016wavenet} autoencoder based unsupervised SVC model is trained to convert among singers appeared in the training set \cite{nachmani2019unsupervised}, where an adversarial speaker classifier is incorporated to disentangle singer information from the encoder output. To further improve this method, PitchNet \cite{deng2020pitchnet} adopts an additional domain fusion term on the pitch to remove pitch information from the encoder output. Variational autoencoder (VAE) \cite{luo2020singing}, generative adversarial network (GAN) \cite{lu2020vaw}, and phonetic posteriorgram (PPG) \cite{li2020ppg} based approaches are also investigated for non-parallel SVC. However, these methods either use acoustic features from conventional vocoders (e.g., WORLD \cite{morise2016world}) or mel spectrograms as intermediate representations during conversion, which may bound the audio quality. 

A very recent unsupervised cross-domain SVC approach (UCD-SVC) \cite{polyak2020unsupervised}, which combines a linguistic extractor with a WaveNet based waveform generator, can convert any source singer to a target speaker/singer appearing in the training set (referred to as any-to-many SVC). UCD-SVC uses a pure convolution network based non-autoregressive model for the waveform-based generator, resulting in very low latency during inference on GPUs. The usage of pitch perceptual loss and automatic speech recognition (ASR) perceptual loss effectively boost the conversion performance. Moreover, UCD-SVC can conduct cross-domain training, i.e., the model can be trained using either speech or singing datasets. However, during conversion UCD-SVC uses three computationally heavy neural networks in the pipeline: the CREPE model \cite{kim2018crepe} for fundamental frequency (F0) computation, Jasper based wave-to-letter acoustic model \cite{li2019jasper} for linguistic feature extraction, and the WaveNet based waveform generator for audio synthesis. This causes the UCD-SVC system to have many parameters, which hinders it from conducting singing voice conversion on CPUs efficiently. Moreover, the training process is complicated and slow.

\begin{figure*}[t]
	\centering
	\includegraphics[width=13cm]{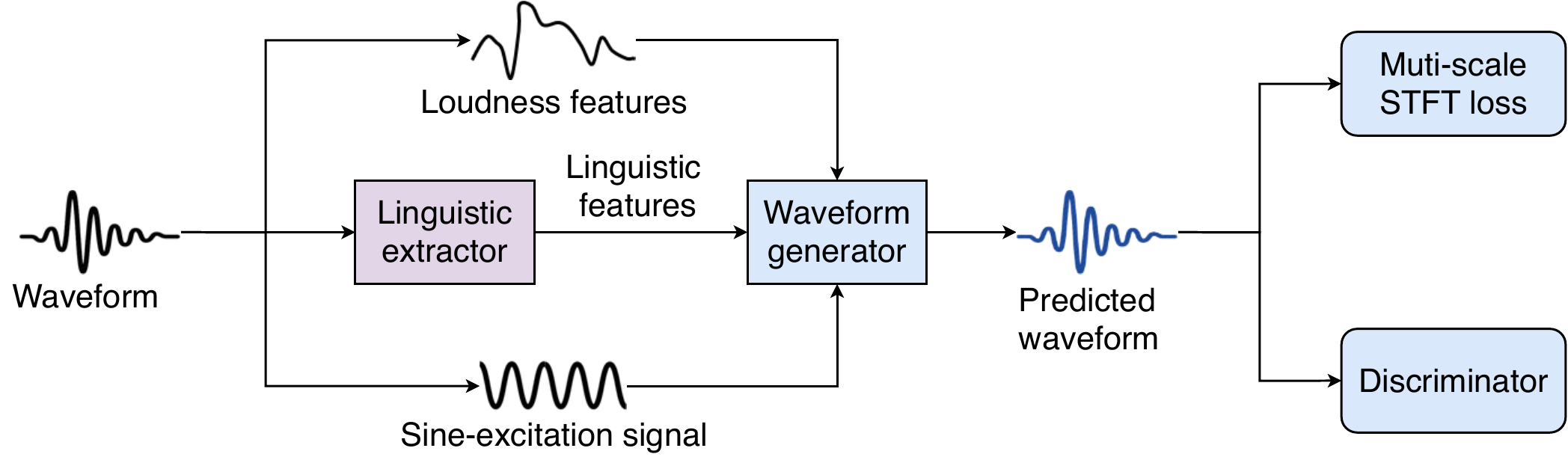}				
	\caption{Schematic diagram of the proposed FastSVC system.}
	\label{sec2:sys_overview}
\end{figure*}

\begin{figure}[t]
	\centering
	\includegraphics[width=8.8cm]{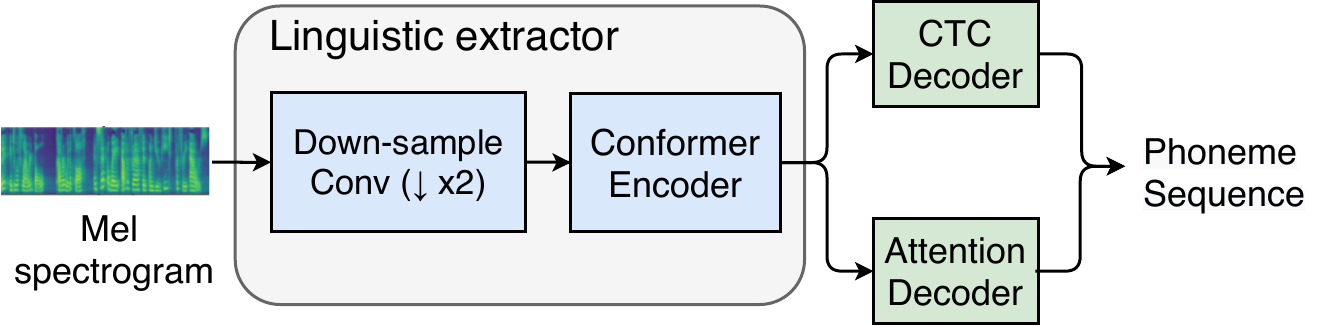}				
	\caption{Hybrid CTC-attention Conformer based model for phoneme recognition.}
	\label{sec2:conformer}
\end{figure}

This paper presents FastSVC, a light-weight cross-domain SVC system, which can achieve high conversion performance in terms of audio fidelity and voice similarity from any source singer, with inference speed 4x faster than real-time on CPUs. The proposed FastSVC approach holds all the merits of UCD-SVC: 1) can conduct cross-domain training; 2) can achieve any-to-many SVC; 3) has desirable singing voice conversion performance. FastSVC takes advantage of recent progress in light-weight end-to-end ASR acoustic modeling, information fusion in deep neural networks and GAN based waveform generative modeling. Specifically, FastSVC uses Conformer \cite{gulati2020conformer} based phoneme recognizer to extract singer-agnostic linguistic features from singing signals. A feature-wise linear modulation (FiLM) \cite{perez2017film} based generator is used to synthesize waveform directly from linguistic features, effectively fusing information from sine-excitation signals and loudness features. The waveform generator can be trained using only a combination of a multi-scale spectral loss and an adversarial loss, which is simpler and faster than the UCD-SVC.


The rest of this paper is organized as follows: Section~\ref{sec2} presents the proposed FastSVC system. Experiments are described in Section~\ref{sec3} and Section~\ref{sec4:conclusion} concludes this paper.

\section{Propsosed Method}
\label{sec2}

The proposed FastSVC system concatenates a linguistic extractor with a waveform generator, as illustrated in Fig.~\ref{sec2:sys_overview}. The linguistic extractor is used to compute singer/speaker-agnostic linguistic features from singing/speech signals, while the waveform generator directly outputs raw waveform from linguistic features, conditioned on sine-excitation signals and loudness features. The linguistic extractor and the waveform generator are trained sequentially, since the waveform generator training process requires linguistic features extracted from a well-trained linguistic extractor. Details of the linguistic extractor and waveform generator are presented in Section~\ref{sec2_1} and Section~\ref{sec2_2}, respectively.

\subsection{Linguistic extractor}
\label{sec2_1}

\begin{figure*}[t]
	\centering
	\includegraphics[width=12cm]{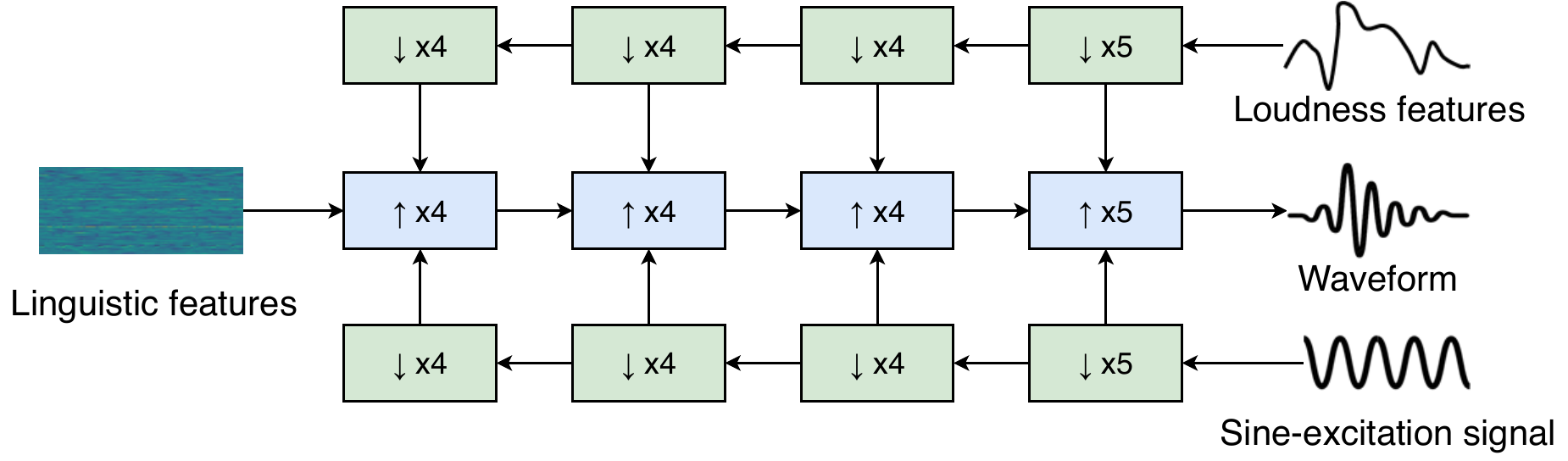}	
	\caption{Schematic diagram of the waveform generator in the proposed FastSVC system.}
	\label{sec2:generator_overview}
\end{figure*}

The UCD-SVC system adopts a pretrained Jasper ASR acoustic model \cite{li2019jasper}, to extract singer/speaker-agnostic linguistic features from wave signals. However, the Jasper model is computationally heavy with more than 300 million parameters. 
One goal of this study is to make the whole SVC system parameter efficient and light-weight, such that inference on modern CPUs can be feasible and fast. The very recent Conformer based ASR acoustic model achieves state-of-the-art recognition performance on LibriSpeech corpus \cite{panayotov2015librispeech}. It is well known that the Transformer models \cite{vaswani2017attention} are good at capturing content-based global interactions, while convolution neural network (CNN) models exploit local features effectively. The Conformer model combines the merits of both the Transformer models and CNN models and also make itself parameter efficient. 

Attracted by the advantages of the Conformer model, this study adopts a Conformer based network structure for the linguistic extractor, as illustrated in Fig.~\ref{sec2:conformer}.
Specifically, we obtain a linguistic extractor from an end-to-end hybrid CTC-attention phoneme recognizer, where the Conformer encoder follows the structure of the small version presented in Table 1 of \cite{gulati2020conformer}. The input spectral features are 80-dimensional log mel spectrograms, on which utterance-level mean-variance normalization is conducted before feeding into the recognizer model. The down-sample layer adopts a strided 2D convolutional structure to down-sample the input mel spetrograms in time scale by a factor of 2, where kernel-size is 4, stride is 2 and output channels is 160. Then the hidden feature maps from the down-sample layer are fed into the Conformer encoder. Then a CTC decoder and attention decoder take the encoder output as input to predict phoneme sequences. The CTC decoder contains one fully-connected (FC) layer. The attention decoder uses location-sensitive attention \cite{chorowski2015attention} mechanism and has one decoder LSTM layer with hidden size of 320. Denote mel spectrogram features as $X$ and phoneme sequence as $Y$, the training loss of the phoneme recognizer is a linear combination of the CTC and attention losses:
\begin{equation} \label{eq1}
\mathcal{L} = \lambda\mathcal{L}_{ctc}(Y|X) + (1-\lambda)\mathcal{L}_{att}(Y|X)
\end{equation}
where $\lambda \in [0,1]$ is a hyper-parameter weighting the CTC objective $\mathcal{L}_{ctc}$ and the attention objective $\mathcal{L}_{att}$. In this paper, we set $\lambda$ to be $0.5$.

The phoneme recognizer is trained with the LibriSpeech corpus (960 hours). After training, we drop the CTC module and attention decoder from the phoneme recognizer and use the remaining part as the linguistic  extractor, which only contains 9 million parameters.

\subsection{Waveform generator}
\label{sec2_2}

As shown in Fig.~\ref{sec2:sys_overview}, the waveform generator takes linguistic features, sine-excitation signals and loudness features as input, and synthesizes waveform directly. 
Linguistic features provide important pronunciation traits for singing voice generation, while sine-excitation signals are melody presentations which proved to be better than F0 contours for SVC tasks \cite{polyak2020unsupervised}. Loudness features make better energy rendering possible in the generated waveform.
An overview of the novel generator model structure is illustrated in Fig.~\ref{sec2:generator_overview}, where the blocks with $\uparrow$ and $\downarrow$ means up-sample blocks and down-sample blocks respectively. The intuitions behind this using two U-shape branches is to fully fuse information from the sine-excitation signals and loudness features into the waveform generation process in different time scales.

\subsubsection{Sine-excitation signals and loudness features}
Sine-excitation signals are computed from the F0 values. Following the NSF models \cite{wang2019neural}, F0 values in frame-rate are first upsampled by linear interpolation to audio-rate and then are regarded as instantaneous frequencies. In voiced segments, the excitation signal are presented as sine waveform, while in unvoiced regions, the excitation signal is represented by Gaussian noise. Denote audio-rate F0 sequence as $\boldsymbol{f}_{1:T}$, following \cite{wang2019neural}, a sine-excitation signal $\boldsymbol{e}_{1:T}$ is computed as:
\begin{equation}\label{eq2}
    \boldsymbol{e}_t =
  \begin{cases}
    0.1 \sin{(\sum_{k=1}^{t}2\pi \frac{f_k}{f_s} + \phi}) + n_t & \quad \text{if } f_t > 0 \\
    100n_t  & \quad \text{if } f_t = 0
  \end{cases}
\end{equation}
where $n_t\sim\mathcal{N}(0,0.003^2)$, $\phi\in[-\pi,\pi]$ is a random initial phase and $f_s$ is the waveform sampling rate.

A-weighting mechanism of the power spectrum, which puts greater emphasis on higher frequencies, is adopted to compute loudness features in this paper. The computation process is identical to that as shown in \cite{hantrakul2019fast}. This paper uses a hop-size of 64 when computing loudness features, which are up-sampled to audio-rate using linear interpolation operation before being fed into the generator.

\subsubsection{Generator model details}

\begin{figure*}[t]
	\centering
	\includegraphics[width=15cm]{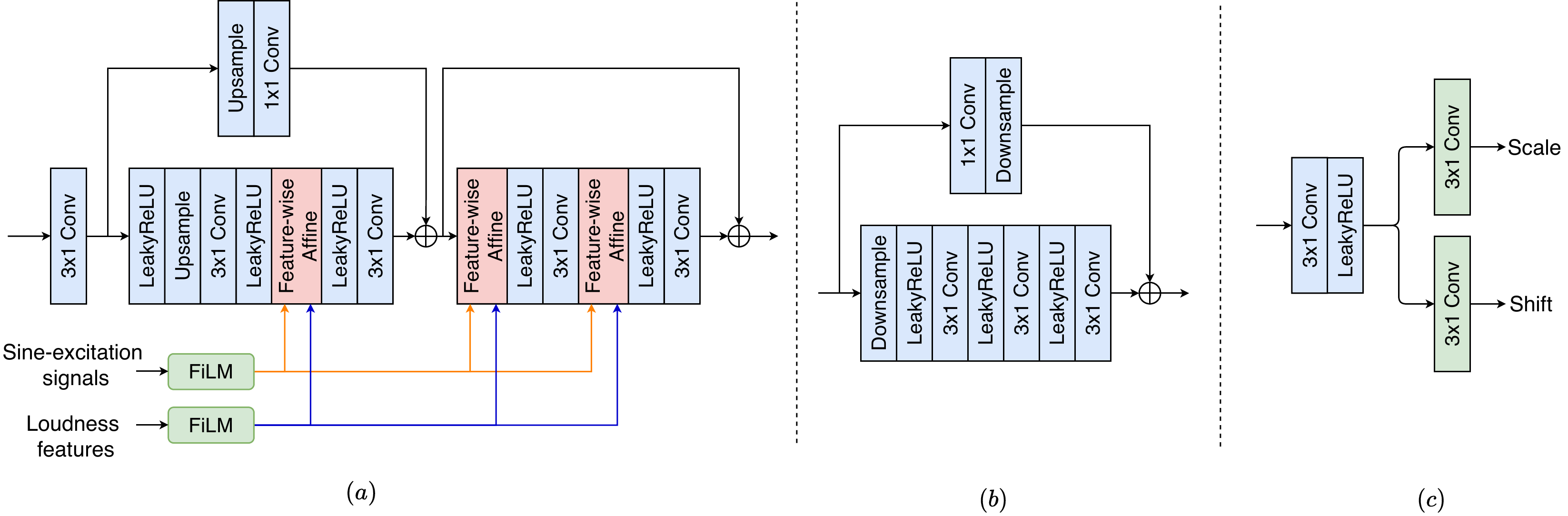} 
	\caption{Network details of building blocks of the waveform generator in the proposed FastSVC system. $(a)$ Network details of the up-sample block. $(b)$ Network details of the down-sample block. $(c)$ The FiLM block appeared in $(a)$. } 
	\label{sec2:model_details}
\end{figure*}

The up-sample blocks and down-sample blocks as shown in Fig.~\ref{sec2:generator_overview} can adopt arbitrary convolutional network structure. The model architecture of the building blocks in the generator is adapted from \cite{chen2020wavegrad} and proper modifications are made for the SVC tasks. Details of the network structure of building blocks in the waveform generator is illustrated in Fig.~\ref{sec2:model_details}.
To convert linguistic features with hop-size of 320 samples into waveform, four up-sample blocks are applied to gradually upsample the temporal dimension by factors of 4, 4, 4, 5 with the number of channels of 192, 96, 48, 24 respectively. The dilation rates are 1, 3, 9, 27 in all up-sample blocks. Down-sample blocks downsample the time dimension of the audio-rate sine-excitation signals and loudness features. The number of channels of the down-sample blocks matches the number of up-sample blocks correspondingly. The dilation rates are 1, 2, 4 in all down-sample blocks. LeakyReLU activation function uses a negative slope of 0.2.

The feature-wise linear modulation (FiLM) \cite{perez2017film} module is used to fuse information from sine-excitation signals and loudness features with the linguistic features, which produces scale and shift vectors given inputs as shown in Fig.\ref{sec2:model_details}$(c)$. FiLM modules have the same number of convolution channels as their corresponding up-sample blocks. The feature-wise affine operation as shown in Fig.\ref{sec2:model_details}$(a)$ is conducted as
\begin{equation} \label{eq3}
    (\gamma_{sine} + \gamma_{loudness}) \odot U_{linguistic} + \xi_{sine} + \xi_{loudness},
\end{equation}
where $\gamma$'s and $\xi$'s represent the scale and shift vectors from the FiLM modules, $U_{linguistic}$ is the up-sampled linguistic features and $\odot$ denotes the Hadamard product. 

In multi-speaker/singer SVC models, the waveform generator has an additional speaker/singer embedding table. The speaker/singer identity information is fused into the up-sample blocks by adding the results from each feature-wise affine operation with speaker/singer embedding vectors. To remove speaker/singer identity information from the results of each feature-wise affine operation, instance normalization \cite{ulyanov2016instance} without affine transformation is performed before combination with speaker/singer embedding vectors.

\subsubsection{Training objectives}

Similar to \cite{polyak2020unsupervised}, the waveform generator is trained under the least-squares GAN \cite{mao2017least} setup, with combination of a multi-scale STFT loss \cite{yamamoto2020parallel}. The discriminator module as shown in Fig.~\ref{sec2:generator_overview} adopts the same multi-scale discriminator architecture presented in MelGAN \cite{kumar2019melgan}. In this section, we denote the three sub-discriminators as $D_k$, $\forall k \in [1, 2, 3]$, the groud-truth waveform as $x$ and the reconstructed waveform as $\hat{x}$.  The generator's adversarial loss $\mathcal{L}_{adv}$ is:
\begin{equation}
    \mathcal{L}_{adv} = \frac{1}{k} \sum_k ||1 - D_k(\hat{x})||_2. 
\end{equation}
The discriminator loss $\mathcal{L}_D$ is computed as:
\begin{equation}
    \mathcal{L}_D = \frac{1}{k} \sum_k(||1 - D_k(x)||_2 + ||D_k(\hat{x})||_2). 
\end{equation}
The multi-scale STFT loss is computed as:
\begin{equation}
    \mathcal{L}_{stft} = \frac{1}{|M|}\sum_{m\in M} (\frac{||S_m-\hat{S}_m||_2}{||S_m||_2} + \frac{||\log{S_m} - \log{\hat{S}_m}||_1}{N}),
\end{equation}
where $S_m$ and $\hat{S}_m$ are the STFT magnitudes computed from $x$ and $\hat{x}$ respectively, with FFT sizes of $m \in M = [2048,1024,512,256,128,64]$ and with 75\% overlap. $N$ is the number of elements. 

The final waveform generator loss $\mathcal{L}_G$ is a linear combination of the adversarial loss and the multi-scale STFT loss as:
\begin{equation}
    \mathcal{L}_{G} = \mathcal{L}_{stft} + \alpha \mathcal{L}_{adv},
\end{equation}
where in this paper, $\alpha=2.5$.

\section{Experiments}
\label{sec3}

\subsection{Experimental setup}
\label{exp_setup}
We choose the UCD-SVC system as the baseline. Both cross-domain and in-domain SVC performance are compared between the UCD-SVC system and the proposed FastSVC system. We also report their cross-lingual SVC performance \cite{sun2016personalized}. 

Three open-source English datasets are used, which are the LJ-Speech corpus \cite{ljspeech17}, the VCTK corpus \cite{yamagishi2019cstr} and the NUS-48E corpus \cite{duan2013nus}. All audio is resampled to 16kHz with mono channel. Datasets are randomly split into train-validation-test sets according to a 90\%-5\%-5\% partition. 
We compare the any-to-one cross-domain (\textbf{A2O-CD}) SVC performance of the UCD-SVC and FastSVC systems by training the models with the single-speaker LJ-Speech corpus, and their any-to-many cross-domain (\textbf{A2M-CD}) SVC performance by training the models with the multi-speaker VCTK corpus (108 speakers in total). Any-to-many in-domain (\textbf{A2M-ID}) and cross-lingual (\textbf{CL}) SVC performance of the UCD-SVC and FastSVC systems are examined by training the models with the multi-singer NUS-48E corpus (12 singers in total). During conversion, source signals are chosen from the NUS-48E test set, except that we use internal Chinese source samples when conducting cross-lingual SVC since there is no open-source Chinese singing dataset.

All models in the UCD-SVC and FastSVC systems are trained at least 600k steps until their losses converge, with batches of 32 one-second long audio segments. The ADAM optimizer \cite{kingma2014adam} with a learning rate of 0.001 is used, where learning rate decays by 0.5 every 100k steps. The discriminator joins the training process after 100k steps. F0 values are extracted using the WORLD vocoder \cite{morise2016world} in the FastSVC system.

\begin{table}[t!]
\caption{Mean opinion score (MOS) results.}
\resizebox{0.48\textwidth}{!}{\begin{tabular}{ccccc}
\thickhline
\multirow{2}{*}{Scenario} & \multicolumn{2}{c}{Naturalness} & \multicolumn{2}{c}{Similarity} \\ \cline{2-5} 
 & UCD-SVC & FastSVC & UCD-SVC & FastSVC \\ \thickhline
A2O-CD & 3.94$\pm$0.09 & 4.00$\pm$0.08 & 3.35$\pm$0.23 & 3.56$\pm$0.20 \\ \hline
A2M-CD & 3.06$\pm$0.08 & 3.52$\pm$0.10 & 2.67$\pm$0.18 & 3.27$\pm$0.22 \\ \hline
A2M-ID & 3.47$\pm$0.07 & 3.48$\pm$0.06 & 3.17$\pm$0.15 & 3.58$\pm$0.19 \\ \hline
CL & 2.92$\pm$0.07 & 3.09$\pm$0.08 & 3.15$\pm$0.19 & 3.26$\pm$0.19 \\ \hline
Recording & \multicolumn{2}{c}{4.62$\pm$0.16} & \multicolumn{2}{c}{-} \\ \thickhline
\end{tabular}}
\label{table:mos_results}
\vspace{-0.5cm}
\end{table}

\begin{table}[t!]
\centering
\caption{Voice similarity.}
\begin{tabular}{cccc}
\thickhline
\multirow{2}{*}{Scenario} & \multirow{2}{*}{Source-target} & \multicolumn{2}{c}{Converted-target} \\ \cline{3-4} 
 &  & UCD-SVC & FastSVC \\ \thickhline
\multicolumn{1}{c}{A2O-CD} & 0.082 & 0.555 & \textbf{0.676} \\ \hline
\multicolumn{1}{c}{A2M-CD} & 0.124 & \textbf{0.588} & 0.433 \\ \hline
\multicolumn{1}{c}{A2M-ID} & 0.234 & 0.712 & \textbf{0.785} \\ \hline
\multicolumn{1}{c}{CL} & 0.274 & \textbf{0.821} & 0.801 \\ \thickhline
\end{tabular}
\label{table:similarity}
\vspace{-0.5cm}
\end{table}

\vspace{-0.3cm}
\subsection{Subjective evaluation}
\label{sub_eval}
Subjective evaluation in terms of both the naturalness and voice similarity of the converted singing samples are conducted\footnote{Audio demo: \url{https://nobody996.github.io/FastSVC/}.}. 
The standard 5-scale mean opinion score (MOS) test is adopted. In the MOS tests for evaluating naturalness, each group of stimuli contains recording samples which are randomly shuffled with the samples converted by the UCD-SVC and FastSVC systems before presented to raters. In MOS voice similarity tests, converted samples are directly compared with the target singers' reference recordings. At least 24 samples are rated for the compared systems in each conversion scenario. We invite 20 Chinese speakers who are also proficient in English to participate in the MOS tests.

The subjective results are presented in Table~\ref{table:mos_results}. 
We can see that in all SVC scenarios, the FastSVC achieves better voice similarity. In terms of audio naturalness, the FastSVC achieves comparable conversion performance to the UCD-SVC system in the any-to-one cross-domain (A2O-CD) and any-to-many in-domain (A2M-ID) scenarios. In the any-to-many cross-domain (A2M-CD) and cross-lingual (CL) scenarios, the FastSVC achieves significantly better performance than the UCD-SVC. The subjective results verify the efficacy of the network structure design and training loss selection in the FastSVC system. 

\subsection{Objective evaluation}
\label{obj_eval}

While MOS is a desired measure for audio naturalness/fidelity in singing voice synthesis tasks, voice similarity is more difficult to subjectively measured since human perception of voice similarity can vary when a singer/speaker utters the same content with different pitch patterns. This can be reflected in Table~\ref{table:mos_results}, where the standard variances of voice similarity MOS values are much bigger than those of naturalness MOS values.

Therefore, we adopt a pre-trained end-to-end speaker recognition model named RawNet2 \cite{jung2020improved} to objectively measure the voice similarity of the converted singing samples. We measure cosine similarities between embedding vectors of audio samples and the desired target speaker embedding vectors before and after conversion, where all embedding vectors are computed by the RawNet2 and singer/speaker embedding vectors are obtained by averaging his/her training audio samples. The voice similarity results are illustrated in Table~\ref{table:similarity}. We can see that both the UCD-SVC and the FastSVC systems can significantly improve cosine similarity of audio sample to a desired target singer/speaker after conversion. 
It is worthy to note that these objective results are not consistent with the subjective results in Table 1, one possible reason is that the pre-trained RawNet2 model is trained using only speech data. It should be better to train a RawNet2 model for speaker/singer embedding vector computation; but we can not access to a large multi-singer corpus during the submission of this paper, this is to be solved in the future work.

\subsection{Inference speed}
\label{infer_speed}

\begin{table}[t!]
\centering
\caption{Inference speed comparison between the UCD-SVC and FastSVC systems. Pytorch implementation without hardware optimization for an Nvidia Tesla P40 GPU and Intel(R) Xeon(R) E5-2680(v4) CPU @ 2.40GHz.}
\resizebox{0.48\textwidth}{!}{\begin{tabular}{cccc}
\thickhline
\begin{tabular}[c]{@{}c@{}}
System \end{tabular} & Parameters & RTF (GPU) & RTF (CPU) \\ 
\thickhline
UCD-SVC & 368.4M & 0.103 & 17.5 \\ 
\hline 
FastSVC (Ours) & \textbf{11.9}M  & \textbf{0.031} & \textbf{0.248}  \\
\thickhline
\end{tabular}}
\label{table:infer_speed}
\vspace{-0.5cm}
\end{table}

The inference speed benchmark results of the compared UCD-SVC and FastSVC systems on both GPUs and CPUs are presented in Table~\ref{table:infer_speed}. All models are implemented with the Pytorch toolkit without any hardware optimization. The proposed FastSVC system has much less number of parameters (11.9M) than the UCD-SVC system (368.4M). Inference speed of the FastSVC system is 3x faster on GPUs and 70x faster on CPUs than the UCD-SVC system. The proposed FastSVC system achieves a real-time factor (RTF) of 0.248 (i.e., 4x faster than real-time) on modern CPUs.

\section{Conclusions}
\label{sec4:conclusion}
In this paper, we have presented FastSVC, a parameter efficient and light-weight cross-domain SVC system, which can achieve superior conversion performance in terms of audio naturalness and voice similarity. The inference speed of the FastSVC system is very fast in both GPUs and CPUs (with real-time factors (RTFs) of 0.031 and 0.248, respectively), which means that FastSVC can be deployed for low-latency real-world applications. Future work include further reducing the parameter size of the FastSVC system and investigating its singer adaptation behavior in the low-resource scenario.

\section{Acknowledgements}
This project is partially supported by a grant from the HKSAR Government's Research Grants Council General Research Fund (Project no. 14208817).
\small
\bibliographystyle{IEEEbib}
\bibliography{icme2021template}

\end{document}